\documentclass[twocolumn,showpacs,preprintnumbers,amsmath,amssymb,prl]{revtex4}

\usepackage{graphicx}
\usepackage{dcolumn,amssymb,amsmath}
\usepackage{bm}

\begin{document}


\title{Observation of wave turbulence in vibrating plates}

\author{Arezki Boudaoud}
\affiliation{%
Laboratoire de Physique Statistique, CNRS / ENS / Paris 6 / Paris 7, 
24 rue Lhomond, 75231 Paris Cedex 5, France
}%

\author{Olivier Cadot}
\author{Beno\^{\i}t Odille}
\author{Cyril Touz\'e}
\affiliation{%
ENSTA-UME, Unit\'e de Recherche en M\'ecanique, Chemin de la Huni\`ere, 91761 Palaiseau, Cedex, France
}%

\date{\today}

\begin{abstract}
The nonlinear interaction of waves in a driven medium may lead to wave turbulence, a state such that energy is transferred from large to small lengthscales. Here, wave turbulence is observed in experiments on a vibrating plate. The frequency power spectra of the normal velocity of the plate may be rescaled on a single curve, with power-law behaviors that are incompatible with the weak turbulence theory of  D{\"u}ring et al. [Phys. Rev. Lett. {\bf 97}, 025503 (2006)]. Alternative scenarios are suggested to account for this discrepancy --- in particular the occurrence of wave breaking at high frequencies. Finally, the statistics of velocity increments do not display an intermittent behavior.
\end{abstract}


\maketitle

The statistical distribution of energy and energy fluxes are central questions concerning out-of-equilibrium dissipative systems with a large number of degrees of freedom. When waves propagate in a medium, their nonlinear interaction might generate other waves with different wavenumbers, which means that energy is transferred between different lengthscales. If the amplitude of waves is large enough, this transfer leads to a distribution of energy on a large number of wavelengths,  and the system reaches a state called wave turbulence~\cite{zakharov}, such that the energy cascades between scales and might be dissipated on a small scale. Although they share the same phenomenology, wave turbulence is much more advanced analytically~\cite{zakharov} than hydrodynamic turbulence~\cite{frisch}. For waves of small amplitude, the framework of weak turbulence yields kinetic equations, the solutions of which have been derived starting from the mid-1960s and correspond to energy spectra with power-law dependence on the wavenumber. Wave turbulence might apply to capillary~\cite{zakharov67a,pushkarev96} or gravity~\cite{zakharov67b,onorato02} waves on the surface of liquids, to plasmas~\cite{musher95}, to nonlinear optics~\cite{dyachenko92}, to magnetohydrodynamics~\cite{nazarenko01} or even to Bose-Einstein condensates~\cite{lvov03}. 

Experimental studies are much less numerous than theoretical ones; they were performed either on the oceanographic scale --- waves on a stormy sea (e.g.~\cite{donelan85}), or on the laboratory scale --- capillary and gravity waves~\cite{holt96,wright97,henry00,brazhnikov02,falcon07a,falcon07b,denissenko07}. Besides, the domain of validity of weak turbulence theory is still a matter of debate. On the one hand, discontinuities in the slope of breaking waves result mathematically in a wide energy spectrum~\cite{phillips85,kuznetsov05}, as apparently observed for gravity waves~\cite{denissenko07}. On the other hand, weak turbulence theory results in Gaussian statistics for the waves, in contrast with experiments when bursts of intense motion occur~\cite{wright97,falcon07b}, a phenomenon known as intermittency. In this context, the theoretical study in~\cite{during06} is very useful as it provides a new system, vibrating plates, where wave turbulence could be observed.

Here we study experimentally a suspended plate driven at high amplitudes~\cite{boudaoud07}. We show that a wide energy spectrum is generated, discuss its interpretation in terms of weak turbulence and wave breaking, and investigate whether the system is intermittent. The typical broadband spectrum observed is also of special interest for its applications, {\em e.g.} for reproducing the sound of thunder in theaters. It is also related to the bright shimmering sound of gongs and cymbals \cite{touze00,CHAIGNE:AcJap:2005}. Transition to chaotic vibration was studied for cymbals in \cite{touze00}, and for panels in \cite{nagai07,amabili05}. 

\begin{figure}
\begin{center}
\includegraphics[width=.4\textwidth]{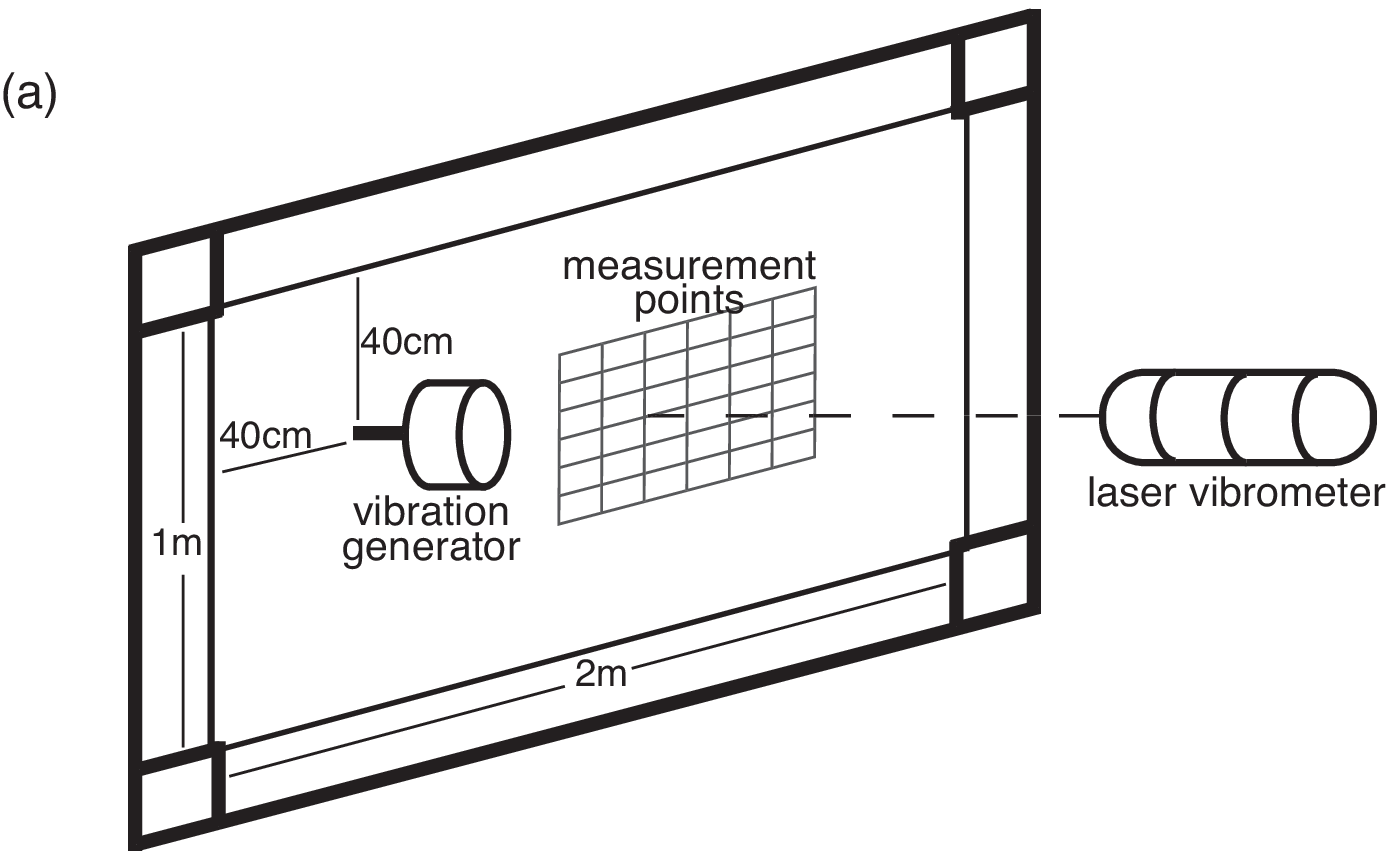}
\vspace{-.2cm}\includegraphics[width=.43\textwidth]{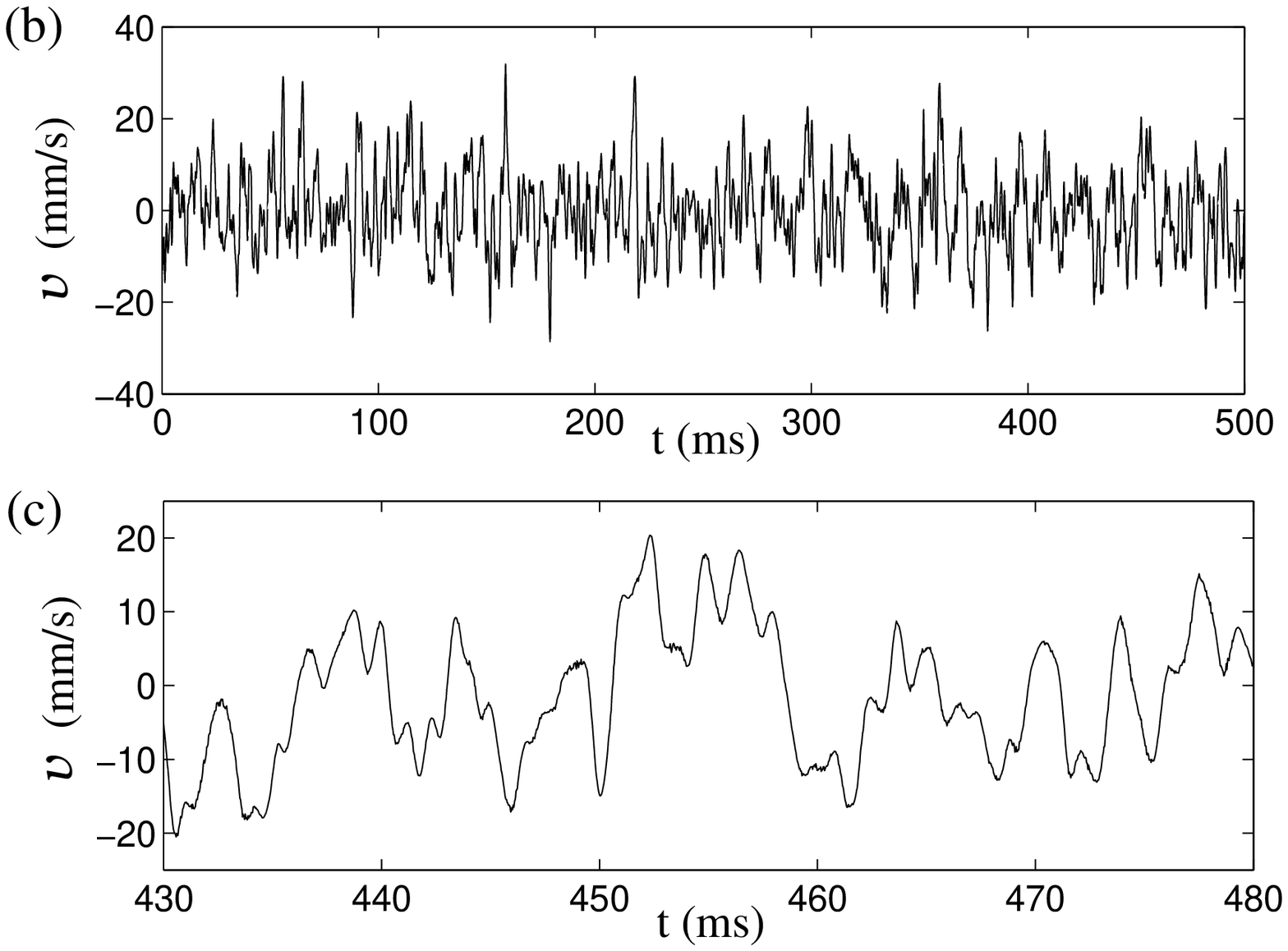}
\caption{The experiment. (a): set-up with a steel plate of dimensions $2\;\textrm{m}\times 1\;\textrm{m}$ and thickness  $h=0.5$ mm; close-up view of the fixation. (b),(c): time series of the local transverse velocity measurements $v(t)$  for the forcing frequency $f_i=20$ Hz;
duration of 10 times (b) and 1 time (c) the forcing period.}\label{fig1}
\end{center}
\end{figure}

The experimental setup consists of a steel plate suspended to a rigid frame and forced with a vibration generator (shaker B{\&}K4810,  glued to the plate with beeswax) moving perpendicularly to the plate ({\sc fig}.~\ref{fig1}a). The plate comes from a reverberation unit named EMT140, that was widely used in studio recordings to add a reverberated sound effect to dry signals recorded by near-field microphones \cite{arcas07}. Hence, the plate was chosen for its very high modal density, obtained by large dimensions $2\;\textrm{m}\times 1\;\textrm{m}$ for a thickness of $h=0.5$ mm, as well as for the moderate values of the quality factor, in order to get a fuzzy reverberated sound. Material properties were estimated as: Young's modulus $E=200$ GPa, Poisson's ratio $\nu=0.3$ and mass per unit volume $\rho=7800\, \textrm{kg/m}^3$. The plate is fixed at its four corners, so that the boundary condition is mainly free. 
The forcing is sinusoidal at $f_\mathrm{i}=20$ Hz that is close to a resonant frequency of the plate; it was chosen in order to allow the best injection of energy in the system, so that the turbulent regime is reached more easily. 
A laser vibrometer gives the normal velocity $v(t)$ at a given point in the plate. The signal is acquired at the sampling frequency of 32~kHz, and the FFT is computed from 50~s of signal, averaged over time windows of 0.5~s, so that $\Delta f$~=~2~Hz. 
A force sensor (impedance head B{\&}K 8001) is mounted between the shaker and the plate. The simultaneous measurement of the velocity at the same point gives the average power $I=\langle Fv\rangle$ injected by the generator into the system (with 1 mW of accuracy).

For a bending wave of frequency $f$ and wavenumber $k$, the dispersion relation is 
\begin{equation}
\label{disp}
f=hck^2  \textrm{, with } c=\sqrt{E/12\rho(1-\nu^2)}/\,2\pi 
\end{equation}
proportional to the sound velocity in the bulk material. It was checked in~\cite{arcas07} that this dispersion relation indeed holds in the present setup. It gives the space-time correspondence of the statistical properties of the velocity signal, similarly to Taylor's hypothesis for fully turbulent flows~\cite{frisch}, when fluctuations are not too large.

\begin{figure}
\begin{center}
\hspace{-.5cm}\includegraphics[width=.38\textwidth]{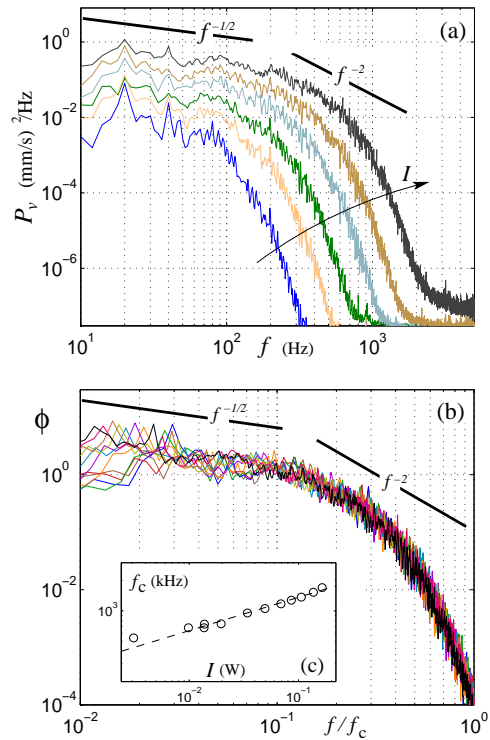}
\caption{Power spectra of the transverse velocity. (a): raw $P_v(f)$ as a function of frequency $f$, for different values of injected power $I$ (in increasing order as displayed by the arrow: $<$~1~mW, 2.3, 8.8, 26.4, 68.8, 136 mW); errors $\Delta f=2$Hz and $\Delta P_v=10^{-7}$ (mm/s$)^2$/Hz. (b): rescaled spectra $(I_0/I)\, P_v$ according to Eq.\ref{scaling} vs. ${f}/{f_c}$ for all forcing amplitudes, where $f_c$ is defined by $(I_0/I)^{1/2}\, P_v(f_c)=10^{-5}$ mm$^2/$s.  Inset (c): evolution of $f_c$  with the forcing intensity . The continuous line is the best power law given in Eq. (\ref{scaling}), yielding an exponent $\alpha=0.33$.}\label{fig2}
\end{center}
\end{figure}

For very low forcing amplitude, the velocity signal $v(t)$ recorded by the vibrometer is sinusoidal. For higher amplitude, it becomes chaotic ({\sc fig}.~\ref{fig1}b,c). In the frequency space, $v(t)$ is characterized by its power spectrum $P_v(f)$, given by the Fourier transform of the auto-correlation function, $P_v(f)=\int \langle v(t)v(t+\tau)\rangle \exp(2\pi i f \tau)\,\mathrm{d}\tau$. This spectrum becomes broadband at high forcing ({\sc fig}.~\ref{fig2}a), which is typical of wave turbulence; however, even with a long time-averaging of the signal, $P_v$ keeps a number of peaks corresponding to the plate eigenfrequencies. We checked homogeneity (by changing the excitation and measurement points) and independence on boundary conditions (by imposing fixed displacements at points at the edge); these changes affected very slightly the power spectra below the injection frequency .

As the forcing amplitude is increased, the spectra exhibit a wider and wider power-law dependence on frequency, $P_v(f) \sim f^{-\beta}$ with $\beta=0.5\pm0.2$ (this error is an upper bound), which would correspond to the cascade regime. It is followed by a fall which could correspond to the dissipative scale. We seek the best rescaling of the spectra as a function of the injected power $I$. This yields the scaling form
\begin{equation}
\label{scaling}
P_v(f)=(I/I_0)^{1/2}\, \phi\left(f/f_c\right)\textrm{,} \quad f_c \propto f_i (I/I_0)^{\alpha}\textrm{.}
\end{equation} 
Here $\phi$ a scaling function, $I_0$ a unit of power and $f_c$ a cut-off frequency. This rescaling enables to collapse the spectra on a single curve ({\sc fig}.~\ref{fig2}b). The exponent for the dependence of $f_c$ on $I$ is found to be $\alpha=0.33\pm0.01$ ({\sc fig}.~\ref{fig2}c). In the cascading frequency range, this implies $P_v(f)\sim I^{1/2+\alpha\beta}f^{-\beta}=I^{0.66\pm0.07}f^{-0.5\pm0.2}$.

In order to compare with previous theoretical work, we first note that the power spectrum for the transverse displacement $\xi$ of the plate is given by $P_\xi(f)\propto P_v(f)/f^2$. When weak turbulence is attained, as investigated in~\cite{during06}, the spatial power spectrum of the displacement can be rewritten as $P_\xi(k) \propto c^{-1}\epsilon^{1/3} k^{-4}$, introducing energy flux  $\epsilon$  per unit mass ($\epsilon$ has units of a velocity cubed and is proportional to the power input $I$), and omitting numerical prefactors and a logarithmic dependence on $k$. $\epsilon$ is proportional to the injected energy $I$. The $1/3$ exponents for $\epsilon$ comes from the $\xi \to -\xi$ symmetry of the plate, which involves four waves interaction. The spectrum can be translated into the frequency space $P_\xi(k)k\mathrm{d}k\propto P_\xi(f)\mathrm{d}f$. Using the dispersion relation~(\ref{disp}), we get $P_\xi(f) \propto h\epsilon^{1/3}f^{-2}$ and $P_v(f)\propto h\epsilon^{1/3}$ is constant. This dependence is significantly weaker than in the measurements ({\sc fig}.~\ref{fig2}a,b).

\begin{figure}
\begin{center}
\vspace{-.2cm}\includegraphics[width=.42\textwidth]{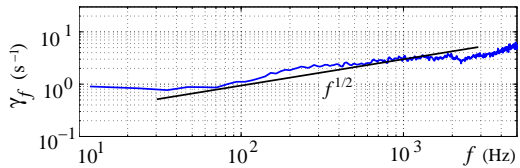}
\caption{Damping factor for the present plate, from \cite{arcas07}.}\label{zedamping}
\end{center}
\end{figure}

In the framework of weak turbulence, nonlinearities of order $p$ imply that $P_v(f)$ scales as $\epsilon^{1/p}$~\cite{zakharov}; the exponent $1/2$ obtained for $p=2$ is the closest to the measured $.66\pm0.07$~(\ref{scaling}). This value of $p=2$ means three waves interactions, a quadratic nonlinearity and no $\xi\to-\xi$ symmetry. Indeed, geometrical imperfections are unavoidable in real plates, which is known to break this symmetry and to produce quadratic nonlinearities~\cite{HuiLeissa83}. Therefore, we assume in the following that $P_v(f)\sim \epsilon^{1/2}f^{-1/2}$, corresponding to a displacement spectrum $P_\xi(f)\sim \epsilon^{1/2}f^{-5/2}$. This assumption allows to investigate the possible role of damping in setting the cutoff frequency. 

Indeed, we introduce the damping rate $\gamma(f)\sim f^{\delta}$. The spectrum of the energy per unit mass is $\mathcal{E}(f)\propto P_v(f)\sim \epsilon^{1/2}f^{-1/2}$. Let us consider the balance of energy over the cascade frequency range; the in-flux is $\epsilon$ while the energy dissipated till $f_c=hc k_c$  is $\int^{k_c} \gamma(k) \mathcal{E}(k)\,\mathrm{d}k\propto\int^{f_c} \gamma(f) \mathcal{E}(f)\sqrt{hc/f}\,\mathrm{d}f$. Balancing these two fluxes  yields $f_c\sim \epsilon^{1/\,2\delta}$. For our setup, a fit to the damping coefficient measured in~\cite{arcas07} is shown in {\sc fig}.~\ref{zedamping} in the frequency domain of interest. It yields $\delta\simeq1/2$, so that $f_c\sim \epsilon\sim I$, which is far from measurements~(\ref{scaling}) and so damping cannot account for the cutoff.

A last option is that the wide energy spectrum might be generated by singularities of the plate displacement as for gravity waves~\cite{phillips85}.  For plates, wave breaking would be replaced~\cite{during06} by ridges~\cite{lobkovsky95,lobkovsky96} and d-cones~\cite{benamar97,cerda99}. It was shown in~\cite{kuznetsov05} that random independent slope discontinuities result in a spectrum
$P_\xi(f)\propto \nu_s \Gamma^2f^{-4}$, $\nu_s$ being the frequency of occurrence of slope discontinuities and $\Gamma$ the rms velocity impulse at each discontinuity. For the velocity $P_v(f)\propto \nu_s \Gamma^2 f^{-2}$ which compares with the second part of the spectra ({\sc fig}.~\ref{fig2}a,b) over half a decade. Besides, the jump should be given by the typical rms velocity $v_\mathrm{rms}~I^{1/2}$, so that we expect $\Gamma \sim \epsilon^{1/2}$. As a consequence, the whole spectrum could result from a 3-waves interaction for low frequencies, as suggested above, and singularities for higher frequencies. These two spectra match
at a frequency $f=f_c$ such that $\epsilon^{1/2}f^{-1/2}\sim \epsilon f^{-2}$, yielding $f\sim \epsilon^{1/3}$,  which agrees with the scaling (\ref{scaling}) as seen in {\sc fig}.~\ref{fig2}c.

\begin{figure}
\begin{center}
\hspace{-.5cm}\vspace{-.2cm}\includegraphics[width=.39\textwidth]{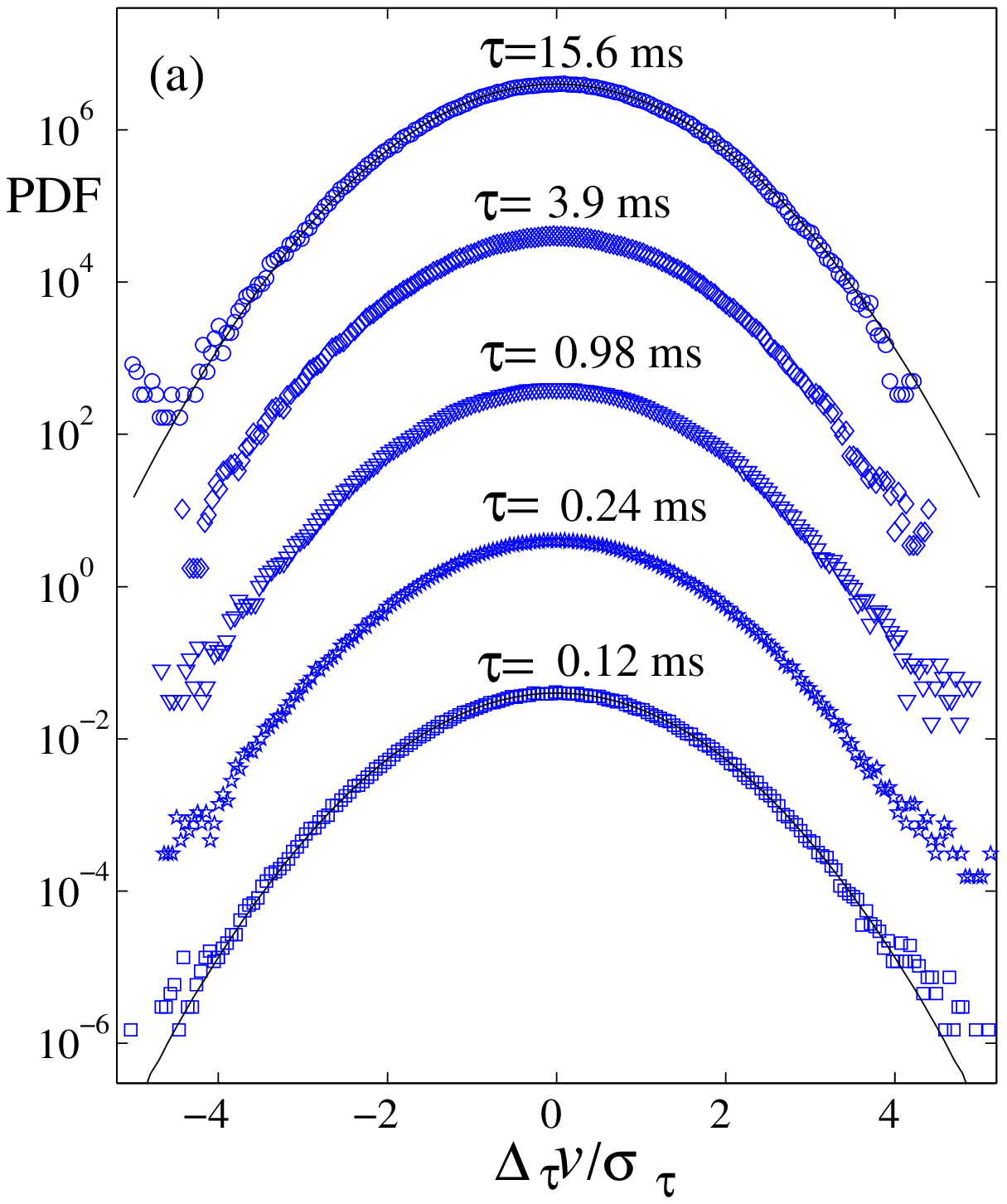}
\vspace{-.35cm}\includegraphics[width=.45\textwidth]{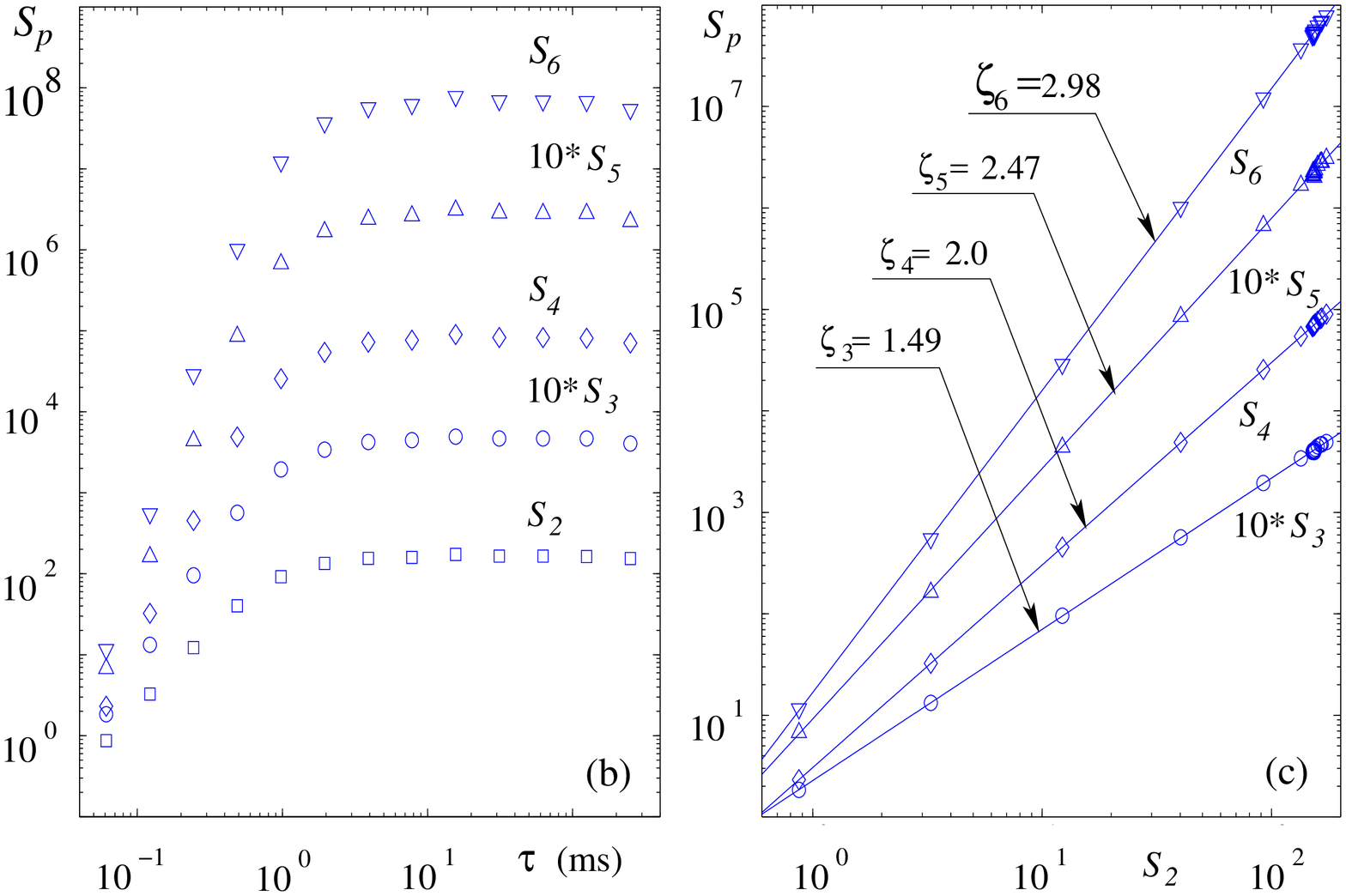}
\caption{Statistical properties of the velocity increments $\Delta_\tau v=v(t+\tau)-v(t)$ (injected power 136 mW). Probability density functions compared to Gaussians in (a). Structure functions of order $p=2, 3, 4, 5, 6$, versus (b): the timelag $\tau$, and (c): the order 2 structure function $S_2$. Continuous lines are best power laws fits with exponents $\zeta_p$ (see text).}\label{fig4}
\end{center}
\end{figure}

Finally, we consider the statistics of  of the velocity increments defined as $\Delta_\tau
v=v(t+\tau)-v(t)$. The PDFs are displayed in {\sc fig}.~\ref{fig4}a for the large forcing
amplitude. An intermittent behavior of the velocity
statistics would be revealed by a change in the PDFs shape
as the lag $\tau$ decreases \cite{frisch}.  Here
we can see in {\sc fig}.~\ref{fig4}a that the PDF shape remains satisfactorily Gaussian
whatever $\tau$. The structure functions,
$S_p(\tau)=\langle\vert\Delta_\tau v\vert^p\rangle$, are plotted in
{\sc fig}.~\ref{fig4}b. They are generally used to determine the
scaling behavior of the velocity differences statistics with the time-lag
$\tau$ \cite{frisch}. The structure functions start to decrease
for $\tau<50$ ms (i.e. the forcing period). For very small
$\tau<0.3$ ms (i.e. the cut-off frequency), the velocity signal
becomes smooth and a simple scaling behavior
$S_p(\tau)=\tau^p$ is found. For wave turbulence, the range of
interest is comprised between these two last extremes. However, within this
range no clear power laws are distinguishable in {\sc fig}.~\ref{fig4}b. We then chose to
plot the structure functions versus $S_2(\tau)$ in
{\sc fig}.~\ref{fig4}c. This technique was used for fully developed
turbulence to measure anomalous scaling exponent due to the
intermittency phenomenon \cite{Benzi}. In our case, the scaling
exponents, defined as: $S_p(\tau) \propto S_2(\tau)^{\zeta_p}$, are indicated in {\sc fig}.~\ref{fig4}c for each order moment $p$. There is no significant deviation from $\zeta_p = p/2$, meaning that
no anomalous scaling is observable. Hence,
wave turbulence in plates does not exhibit any intermittency
phenomenon. 

To summarize, we observed a broadband spectrum in a vibrating plate and investigated the variations of the cut-off frequency. In this context, internal damping mechanisms (mainly thermoelastic and viscoelastic losses for our plate~\cite{arcas07}) seem to be irrelevant. Losses at the edge~\cite{during06} can be discarded as the plate is fixed only at the corners. The radiation of acoustic waves in air is negligible since the frequencies of interest are well below the coincident frequency, for which  bending and acoustic waves have the same phase velocity. The value of  this frequency has been measured as  20 kHz in our set-up~\cite{arcas07}. For thicker plates, the coincident frequency may fall in the frequency range of interest, thus leading to a huge increase of the damping factor, see {\em e.g.}~\cite{chaigne2001}. This could affect the conclusions on the cut-off. Our experimental results suggest a 3-waves spectrum matched to a spectrum of singularities where dissipation occurs. Obviously they call for more theoretical effort, in particular concerning the weak turbulence of plates with quadratic nonlinearities or the turbulence of singularities. 
\begin{acknowledgments}
We are grateful to K. Arcas, E. Hamm, F. Melo and S. Rica for help and discussions. J.-M. Mainguy and L.-C. Tr\'ebuchet from Radio-France are also thanked for the loan of the plate reverberator. This work was partially supported by ANR Blanc OPADETO. 
\end{acknowledgments}

\end{document}